\documentclass[prd,nofootinbib,showpacs,preprintnumbers,amssymb,11pt]{revtex4} % not twocolumn
\usepackage{graphicx}% Include figure files
\usepackage{epsfig}
\usepackage{dcolumn}% Align table columns on decimal point
\usepackage{bm}% bold math
\usepackage{amsmath}
\def\gsim{\lower0.5ex\hbox{$\:\buildrel >\over\sim\:$}}
\def\lsim{\lower0.5ex\hbox{$\:\buildrel <\over\sim\:$}}

\let\d=\delta

\newcommand{\be}{\begin{equation}}
\newcommand{\ee}{\end{equation}}
\newcommand{\bea}{\begin{eqnarray}}
\newcommand{\eea}{\end{eqnarray}}

\newcommand{\del}{\partial}

\newcommand{\nbox}{{\,\lower0.9pt\vbox{\hrule \hbox{\vrule height 0.2 cm
\hskip 0.2 cm \vrule height 0.2 cm}\hrule}\,}}

\begin{document}

\preprint{UCI-TR-2010-14}

\title{On the proper treatment of massless fields in Euclidean de Sitter space}

\author{Arvind Rajaraman}
\email{arajaram@uci.edu}
\affiliation{Department of Physics and Astronomy, University of California,
Irvine, CA 92697, USA}

\date{\today}

\begin{abstract}

We analyze infrared divergences arising in calculations involving light and massless fields
in de Sitter space. We show that these arise from an incorrect treatment of the constant mode
of the field, and show that a correct quantization leads to a well-defined and calculable
perturbation expansion. We illustrate this by computing the first nontrivial loop correction
in a theory of a massless scalar field with a quartic interaction.

\end{abstract}

\pacs{98.80.Cq, 98.80.Qc, 11.10.Ef}

\maketitle

\section{Introduction}

There has been much recent interest in the possibility of probing
nongaussianities in the CMB at the WMAP and PLANCK experiments
~\cite{Komatsu:2008hk, Slosar:2008hx,Bartolo:2004if}.
Such nongaussianities occur in theories where the inflaton has significant self-interactions
(and more generally if it has interactions with other fields
participating in the inflationary epoch).
It is thus of great importance to calculate the prediction for nongaussianities in
a theory with inflaton interactions.
Unfortunately, the existence of light fields leads to infrared divergences
which make such calculations suspect
(see, for example,
\cite{Weinberg:2005vy,weinberg1,Starobinsky:1994bd,
%Allen:1986ta,Higuchi:2001uv, Higuchi:2002sc,Kirsten:1993ug,Folacci:1992xc,
Riotto:2008mv}).

These divergences often arise because of the
infinite expanding volume of de Sitter space. The
volume of the metric grows exponentially in the global
time coordinate, which leads to
an effective growth in the coupling at late times. While this is
compensated to some extent by
the falloff of the wave functions, it can be shown that
even tree
level scattering amplitudes grow with time \cite{weinberg1,marolf,meulensmit}.
In loop diagrams, this same effect leads to late-time divergences
which have been argued
to %In fact it has been argued that these divergences
signal the breakdown of de Sitter field theory~\cite{polyakov1},
perhaps leading to a decay
to another vacuum~\cite{polyakov2}. On the other hand, these divergences can be
eliminated entirely by first taking the
Euclidean continuation of the theory, which transforms
de Sitter space to the Euclidean sphere.
The sphere, being compact,
does not have large volume divergences. Correlation functions are
therefore infrared finite, and can be
continued back to
the Lorentzian theory to obtain well defined results~\cite{marolf}. This
suggests that the late time divergences are
unphysical,
and arise from  the breakdown of the in-out formalism in Lorentzian
de Sitter space in global coordinates~\cite{higuchinew}.
We henceforth assume that the true definition of the
de Sitter theory is by a continuation from the
theory on the Euclidean sphere.

Even though the Euclidean formalism eliminates divergences coming from
the infinite volume of de Sitter space,
light or massless particles can still lead to infrared divergences.
For example, in the theory of a scalar field $\phi$ with
a mass $m$ and a interaction $\lambda \phi^4$,
a $k$-loop diagram has a factor $\left({\lambda H^2\over m^2}\right)^k$~\cite{Burgess:2010dd}.
For $m^2\ll \lambda H^2$,
the loop diagrams are larger than the tree contribution, and perturbation theory seems to
break down. This is a problem for theories of slow roll inflation in which
the inflaton is very light field; gravitons also behave similarly to massless
scalars, and may potentially lead to divergences. Understanding the massless limit is therefore
important for calculations in perturbation theory.

In this note, we
reexamine the
 $\lambda \phi^4$ model in de Sitter space.
 Our main result is that even in the limit
 of very small masses
($m^2\ll \lambda H^2$),
the perturbation theory of this model
is under control.
We show that the $k$-loop diagram scales as
$(\sqrt\lambda)^k$ in the limit $m^2\rightarrow 0$. The apparent infrared divergences thus
change the expansion parameter from $\lambda$ to $\sqrt\lambda$;
nevertheless, for
small $\lambda$, the perturbation theory is well defined.
For the formalism on the Euclidean sphere, these
corrections can be computed explicitly.

In the next section, we begin by reviewing the basic features of de Sitter space,
and discuss how light fields lead to
 the appearance of infrared divergences. We then show that in the Euclidean formalism,
 the divergences are entirely due to the
 incorrect treatment of one mode. We show in section III that the correct treatment
 of this mode removes the divergences and makes tree level correlation functions finite.
 In section IV we extend this to loop diagrams, and show that they are also well defined and calculable;
 %nd yields well defined and calculable results.
we illustrate this by  computing the first nontrivial
loop correction to the two-point
correlation function of the field in the massless theory. We
then discuss how our methods may be modified for
the in-in formalism, and finally end with a discussion of our results.

\section{Scalar Field Theory in de Sitter space}

There are many possible coordinatizations of de Sitter space
~\cite{Spradlin:2001pw}. We shall be concerned
mainly with the coordinatization in global coordinates.
In these coordinates, the metric of $D$-dimensional de Sitter (denoted $dS_{D}$) can be written
%in global coordinates
as
\bea
ds^2=R^2(-dt^2 + \cosh^2t\ d\Omega^2)
\eea
where $d\Omega^2$ is the metric on the $d$-sphere with $D=d+1$. These coordinates cover all of
de Sitter space.

In these coordinates, the volume of de Sitter grows exponentially towards both
past and future infinity. This exponential growth can lead to
infrared divergences in calculations in field theory, since the
effective coupling is scaled by the volume.
A simple example of this can be seen by considering a massive scalar field
theory in de Sitter space~\cite{higuchinew2}.
This theory can be solved exactly, and the modes can be solved for.
On the other hand, one can attempt to treat the mass as a perturbation by  first finding
the propagator of the
massless theory, and finding the corrections to the propagator as a
function of the mass.
Curiously, the corrections are found to be divergent
at each order, even though the answer is known to
be finite. This can be shown to be due to a failure of the
standard in-out formalism of field theory~\cite{higuchinew2}.
%\bea
%G(x,y)=
%\eea

To find a well defined formalism for quantum field theory in de Sitter space,
we perform a Wick rotation of the metric.
After a rotation $t\rightarrow i(\tau-\pi/2)$, the Euclidean metric
is found to be
\bea
ds^2\equiv R^2g_{\mu\nu}dx^\mu dx^\nu=R^2(d\tau^2 + \sin^2\tau d\Omega^2)
\eea
To obtain a regular metric, $\tau$ must be compactified
by the identification $\tau \rightarrow \tau+2\pi$.
%so that
$g_{\mu\nu}$ is then the metric on the $D$-sphere of unit radius.

We will consider a scalar field of mass $m$ with a quartic
interaction in de Sitter space. Our goal will be to calculate correlation functions
in this theory, with particular
emphasis on the case of very small  masses.
The Euclidean action is
\bea
S_E = \int d^D x R^D\sqrt{g}( R^{-2}g^{\mu\nu}\del_\mu\phi\del_\nu\phi+m^2\phi^2+\lambda\phi^4)
\eea
%where $g_{\mu\nu}$ is the metric on the n-sphere of unit radius.
As usual we begin by solving the quadratic part of the action.
The equation of motion is
\bea
\nabla^2\phi+m^2R^2\phi=0
\eea
where $\nabla^2\equiv {1\over \sqrt{g}}\del_\mu \sqrt{g} g^{\mu\nu}\del_\nu$ is the Laplacian
on the unit $D$-dimensional sphere.
This equation can be solved in terms of spherical harmonics on the $D$-dimensional sphere.

The $D$-dimensional spherical harmonics have been discussed by several
authors (see e.g. \cite{Higuchi:1986wu}).
While on the 2-sphere, the harmonics are labeled by two integers $L, m$ with $|m|<L$, the
harmonics on
the $D$-sphere are labeled by a vector of integers
$\vec{L}=(L,L_{d}...L_1)$
%is a n dimensional vector of integers
with
$L\geq L_{d}\geq... L_2\geq |L_1|$.  $L$ is the total angular momentum.
We denote the corresponding spherical harmonics as $Y_{\vec{L}}$. These satisfy
\bea
\nabla^2 Y_{\vec{L}}=-L(L+d) Y_{\vec{L}}
\eea
 These spherical harmonics also satisfy the orthogonality relations
 \bea
 \sum_{\vec{L}}Y_{\vec{L}}(x)Y^*_{\vec{L}}(y)=\sqrt{g}\delta^D(x-y)
 \\
  \int_{S^D} d^Dx \sqrt{g}\ Y_{\vec{L}}(x)Y^*_{\vec{M}}(x)=\delta_{\vec{L}\vec{M}}
  \label{ortho}
 \eea
 Much of our discussion will center around the $\vec{L}=0$ solution,
 which we will call the "zero-mode".%, in an abuse of notation.

 The spherical harmonics form a complete set of states. The field $\phi$ can therefore
  be expanded in modes on the sphere as
 \bea
 \phi=\sum_{\vec{L}}\phi_{\vec{L}}Y_{\vec{L}}
 \eea
 In terms of these modes, the quadratic part of the action becomes
 \bea
 S_2=R^{d-1}\sum_{\vec{L}} (L(L+d)+m^2R^2)|\phi_{\vec{L}}|^2
 \eea

 The exact Euclidean two-point function is defined by the path integral to be
 \bea
 \langle\phi(x) \phi(y)\rangle={\int {\cal D}\phi\ \phi(x) \phi(y)\exp(-S_E(\phi))
 \over \int {\cal D}\phi\ \exp(-S_E(\phi))}
 \eea

 If we replace the action by $S_2$, and carry out the usual procedure of
 path integration over bosonic fields,  we find the Euclidean correlation function
  \bea
 \langle\phi(x) \phi(y)\rangle=\sum_{\vec{L}} {Y_{\vec{L}}(x) Y^*_{\vec{L}}(y)\over R^{d-1}
 ( L(L+d)+m^2R^2)}
 \label{sum1}
 \eea

 The correlation function can also be written
 directly in position space
 as~\cite{marolf}
 \bea
 \langle\phi(x) \phi(y)\rangle={1\over 4\pi^{d/2+1}}{\Gamma(d/2)\Gamma(-\sigma)
\Gamma(d+\sigma)\over \Gamma(d)}\ {}_2F_1\left(-\sigma,d+\sigma;(d+1)/2; {1+Z_{xy}\over 2}\right)
\eea
where $Z_{xy}$ is the chord distance between the points $x,y$ when the sphere is embedded in
a flat $R^{D+1}$, and we have defined
$\sigma=-{d\over 2} + \sqrt{{d^2\over 4}-m^2R^2}$.
%\bea
% Z=
% \eea
The de Sitter correlation function can
be obtained by rotating the Euclidean correlation function to Lorentzian signature with
an appropriate $i\epsilon$ prescription~\cite{marolf}.

As we have mentioned, since the sphere is compact, there
are no divergences arising from integrations over the
volume of the space (in contrast to de Sitter which is noncompact).
All Euclidean correlation functions are hence finite, and
can be continued back to Lorentzian signature to obtain finite results in de Sitter space.
This is however only true if the fields
are massive; when $m^2\rightarrow 0$, we get a divergence both in tree and loop
amplitudes.
At tree level, this is most easily seen by looking at the correlation function.
%As $m^2\rightarrow 0$,
For small masses, we find $\sigma\rightarrow -{m^2\over d}$ and the propagator approaches the limit
\bea
%G(x,x')\rightarrow {1\over \sigma}\label{divgce}
 \langle\phi(x) \phi(y)\rangle\rightarrow-{1\over 4\pi^{d/2+1}}{\Gamma(d/2)
\over \sigma} \label{divgce}
\eea
which is divergent as $m^2\rightarrow 0$.

The same problem will occur in loop diagrams. The one-loop diagram
has both a ultraviolet and an infrared divergence, but since it only contributes to
the renormalization of the mass, it
is completely canceled by the mass counterterm. The leading divergence then comes
from the sunset diagram shown here.
\begin{figure}[h]
\centerline{\includegraphics[width=3 cm]{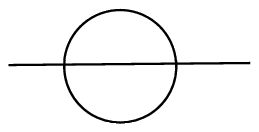}}
%\caption{caption{Plot of allowed $N_1$ vs $N_2$ mass.  The region in the upper
%right hand is allowed. }}%%
%\label{fig:2pgm}
\end{figure}

The internal propagators are divergent in the massless limit, and this induces a divergence in the loop integral.
More generally, a $k$-loop diagram has a factor $({\lambda H^2\over m^2})^k$, and for
$m^2\ll \lambda H^2$, the perturbation theory breaks down.

\section{A proper treatment of the zero-mode}

Looking at the formula for the correlation function,
it is clear that the problem comes from the zero-mode. For zero masses, the
$\vec{L}=0$ term gives a divergent contribution to the sum (\ref{sum1}), which then leads
to the divergence in (\ref{divgce}). In loop diagrams, each zero-mode propagator comes with a factor
of ${1\over m^2}$, and for the maximal number of zero-mode propagators,
we find a scaling $({\lambda\over m^2})^k $ for a $k$-loop diagram.

The problem with the zero-mode may be traced
to the fact that the
terms in the action quadratic in $\phi_{\vec{0}}$ are vanishing when $m=0$.
 For small masses, the fluctuations are proportional to an inverse power of the mass,
 which can be large.
 In the limit of zero masses, fluctuations of the zero-mode are unsuppressed, and the path
 integral
 diverges.

   For the free theory, this statement is exact but uninteresting since the fluctuations cannot be
   observed. However, in the interacting theory,
   %the interaction term can be important for the analysis of the
   %fluctuations;
   it cannot be the case that the fluctuations are unsuppressed; for instance,
   the $\phi^4$ term would limit the
   fluctuations of the field even if the mass was zero.
   Indeed, the large fluctuation of the zero-mode
indicates that the terms which are higher order in the zero-mode cannot be neglected.
The breakdown of perturbation theory in the small mass limit is thus understood
as coming from an incorrect treatment of the zero-mode;
we tried to quantize this mode using the quadratic terms alone, while we should have kept the
interaction terms.

More precisely, we only need to keep the
$\phi_{\vec{0}}^4$
term at leading order; i.e. terms
like $\phi_{\vec{0}}^2\phi_{\vec{L}\neq 0}^2$
representing interactions between the zero-modes
and the nonzero-modes can be treated
as perturbations.
This is roughly because  each of the nonzero-modes has a fluctuation
which is small compared to the fluctuation
of the zero-mode.
The terms involving interactions of only nonzero-modes can also be treated
as perturbations for the same reason. We shall explicitly justify this below.

To make the above statements quantitative, we now proceed to
quantize the theory keeping the $\phi_{\vec{0}}^4$ term.
We will be working in the limit where $m^2\ll \lambda H^2$, which is the
parameter region where the infrared divergences are important.
These masses are parametrically small, and can be treated in
perturbation theory (assuming of course that the $m\rightarrow 0$ limit exists).
We will therefore begin by taking the mass to be zero. We will also henceforth
set $R=1$ by a rescaling of the fields.

The leading order action is then
\bea
S_0&=&\lambda \phi_{\vec{0}}^4 |Y_{\vec{0}}|^2
+\sum_{\vec{L}\neq 0} L(L+d)|\phi_{\vec{L}}|^2
\\
&=&\lambda \phi^4_{\vec{0}}{\Gamma((d+2)/2)\over 2\pi^{(d+2)/2} }
+\sum_{\vec{L}\neq 0} L(L+d)|\phi_{\vec{L}}|^2
\eea
where we used the orthonormality condition (\ref{ortho})
%To find the position space correlation function, we first find
to determine $Y_{\vec{0}}(x)$ (we have also chosen a phase convention where it is real). %which
%is determined
%by  to be %satisfy
%\bea
%|Y_{\vec{0}}(x)|^2{2\pi^{(d+2)/2}\over \Gamma(d+2/2)}=1
%\eea

 Note that this is now not a Gaussian theory, which would normally make
 it impossible to work with.  In fact, in addition to being nonlinear, the
 zero-mode is strongly coupled; $\lambda$
 can be scaled out of the $\phi_{\vec{0}}^4$
 interaction.
 In this case, only one mode is involved in the nonlinear
 interaction. We can therefore hope to solve this single mode exactly as in quantum
 mechanics (indeed it is even simpler than quantum mechanics, since there is no time dependence either).

To illustrate this, we find the exact two point function in this theory.
The two-point function is now
 \bea
 \langle\phi(x) \phi(y)\rangle &=& {\int {\cal D}\phi\ \phi(x) \phi(y)\exp(-S_0(\phi))
 \over \int {\cal D}\phi\ \exp(-S_0(\phi))}
 \\
& = &{\int {\cal D}\phi_{\vec{0}}\ \phi^2_{\vec{0}}\ Y_{\vec{0}}(x) Y_{\vec{0}}(y)
 \exp(-\lambda_{eff}\phi_{\vec{0}}^4)
 \over \int {\cal D}\phi_{\vec{0}}\ \exp(-\lambda_{eff} \phi_{\vec{0}}^4)}
 +\sum_{\vec{L}\neq 0} {Y_{\vec{L}}(x) Y^*_{\vec{L}}(y)\over L(L+d)}
  \\
& = &{c_2Y_{\vec{0}}(x) Y_{\vec{0}}(y)\over \sqrt{\lambda_{eff}}}+\sum_{\vec{L}\neq 0}
 {Y_{\vec{L}}(x) Y^*_{\vec{L}}(y)\over L(L+d)}\label{sum2}
 \eea
where we have introduced the constant
\bea
c_2={\int dx\ x^2\exp(-x^4)
 \over \int dx\exp(-x^4)} = {\Gamma({3\over 4})\over \Gamma({1\over 4})}
 \eea
 and we have defined
 \bea
 \lambda_{eff}=\lambda{\Gamma((d+2)/2)\over 2\pi^{(d+2)/2}}
 \eea

 We thus have an exact solution for the correlation function, which
 is finite even in the massless limit.
 In particular, the $\vec{L}=0$ mode gives a finite contribution to the
 propagator in this
 limit.

 When the mass term is nonzero and small, the correlation function
 is corrected.  This can be done by  treating the mass term as
 a perturbation in the
 action and resumming
 the corrections in the standard way.
The result is to shift the poles in the correlation function,
which now becomes
 \bea
  \langle \phi(x)\phi(y)\rangle={Y_{\vec{0}}(x)
  Y_{\vec{0}}(y)\over {{\sqrt{\lambda_{eff}}\over c_2}+m^2}}
  +\sum_{\vec{L}\neq 0} {Y_{\vec{L}}(x) Y_{\vec{L}}(y)\over L(L+d)+m^2}\label{correctprop}
 \eea

Comparing (\ref{sum1}) and (\ref{sum2}), the
correlation function is position space is now
 \bea
  \langle \phi(x)\phi(y)\rangle
  %G(x,x')
  ={1\over 4\pi^{d/2+1}}{\Gamma(d/2)\Gamma(-\sigma)
\Gamma(d+\sigma)\over \Gamma(d)}\ {}_2F_1(-\sigma,d+\sigma;(d+1)/2; {1+Z_{xy}\over 2})
\nonumber\\
-{\Gamma({d+2\over 2})\over 2\pi^{(d+2)/2}}\left({1\over m^2}-{1\over
{\sqrt{\lambda_{eff}}\over c_2}+m^2}\right)
\eea
which in the limit $m^2\rightarrow 0$ %(in this limit $\sigma = -{M^2\over d}$)
goes to the finite limit
 \bea
  \langle \phi(x)\phi(y)\rangle\rightarrow
 % {\sqrt{\Gamma({d+2\over 2})}\over \sqrt{2}\pi^{(d+2)/4}}
 {c_2\over {\sqrt{\lambda_{eff}} }}
\eea
This is our main result; the proper quantization of the zero-mode has modified the correlation
function in such a way as to render the massless limit finite.

 \section{Calculations in the massless theory}

 We now show explicitly that the proper treatment of the zero-mode leads
  to finite results for tree and loop diagrams.
We shall work in the massless limit; small masses may be treated in a perturbation expansion.

It is immediate that the tree level diagrams are all finite. As we have shown above the
 propagator is finite in the massless limit. Since the sum over the higher harmonics is convergent,
 any product of propagators will yield a finite result.

The finiteness of the loop diagrams is not as clear, and requires more analysis.

Consider the sunset diagram in the previous section.
If the external states are zero-modes ($\vec{L}=0$), there is a contribution to the loop where all
the internal lines are also zero-mode propagators. This contribution has two factors
of $\lambda$ from the vertices, and four factors of ${1\over \sqrt{\lambda}}$ from
the propagators, and is therefore an ${\cal O}$(1) correction to the correlation function.
This is a direct indication that the zero-modes are strongly coupled; the loops are as
important as the leading order diagram. Fortunately, we have already solved this problem; we have
found above the {\it exact} two point correlation function of the zero-modes i.e. the resummation of these
diagrams has already been performed in obtaining the leading
order correlation function (\ref{correctprop}).

We now consider the case where the external legs are not zero-modes
($\vec{L}\neq 0$). The leading effect then comes from diagrams
where two of the internal lines
are  zero-mode propagators (momentum conservation prevents all three from being zero-mode
propagators). This contribution once again has two factors
of $\lambda$ from the vertices, but now there are only 2 factors of ${1\over \sqrt{\lambda}}$ from
the zero-mode propagators. The correction therefore scales as ${\cal O}(\lambda)$,
and is a parametrically small correction to the correlation function.

The interactions
of zero-modes and nonzero-modes can therefore be treated in perturbation theory, confirming
the  argument above.
However, to explicitly
find the ${\cal O}(\lambda)$ correction to the correlation function, we should check whether higher order
diagrams are suppressed. In fact, as we now see, this is not the case.

Higher order diagrams will involve both interactions between
zero-modes and nonzero-modes, which can be treated in
perturbation theory, as well as zero-mode self interactions, which must
be treated exactly.
We show below such a representative diagram which contributes to corrections to a
nonzero-mode correlation function (we have denoted the nonzero-mode correlation function by a thicker line).
\begin{figure}[h]
\centerline{\includegraphics[width=3 cm]{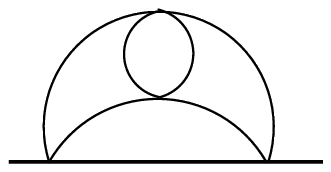}}
%\caption{caption{Plot of allowed $N_1$ vs $N_2$ mass.  The region in the upper
%right hand is allowed. }}%%
%\label{fig:2pgm}
\end{figure}
There are four vertices and 6 zero-mode propagators, so the
diagram scales as ${\cal O}(\lambda)$, which is the same scaling as the
sunset diagram correction. This means that we have to resum all these higher loop
 diagrams to get the
${\cal O}(\lambda)$ correction to the nonzero-mode correlation function.
This is again because the zero-modes are
strongly coupled.

The full set of diagrams that we need to resum involves further
vertices involving only zero-modes. It is easy to see that the full set of such diagrams
is of the form shown in the diagram below.
\begin{figure}[h]
\centerline{\includegraphics[width=3 cm]{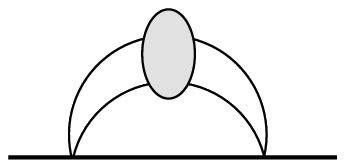}}
%\caption{caption{Plot of allowed $N_1$ vs $N_2$ mass.  The region in the upper
%right hand is allowed. }}%%
%\label{fig:2pgm}
\end{figure}

The blob is any interaction of the zero-modes alone.
While it is impossible to calculate these diagrams perturbatively, we can
write the full set of diagrams as  nonzero-mode propagators
convolved with a four point function of zero-modes
%in the quantummechanics.This diagram is then
\bea
288\lambda^2\int d^3x_1 d^3 x_2 |Y_{\vec{L}}(x_1)|^2% Y^*_{\vec{L}}(x_1)
|Y_{\vec{L}}(x_2)|^2
%Y^*_{\vec{L}}(x_2)
Y^2_{\vec{0}}(x_1) Y^2_{\vec{0}}(x_2)\nonumber
 (\langle\phi_{\vec{L}}\phi^*_{\vec{L}}\rangle)^3
 %\langle\phi_{\vec{L}}
%\phi^*_{\vec{L}}\rangle\langle\phi_{\vec{L}}\phi^*_{\vec{L}}\rangle
\times
\langle\phi_{\vec{0}}^2\phi_{\vec{0}}^2\rangle
\eea

We evaluate
\bea
\langle\phi_{\vec{0}}^4\rangle={\int {\cal D}\phi_{\vec{0}}
\ \phi_{\vec{0}}^4
\exp(-\lambda_{eff} \phi_{\vec{0}}^4)
 \over \int {\cal D}\phi_{\vec{0}}\ \exp(-\lambda_{eff}  \phi_{\vec{0}}^4)}
 ={c_4\over \lambda_{eff}}
\eea
where
\bea
c_4={\int dx\ x^4\exp(-x^4)
 \over \int dx\ \exp(-x^4)}={1\over 4}
 \eea
The final correction to the correlation function is then of order $\lambda$.
The corrected correlation function is
\bea
 \langle\phi(x) \phi(y)\rangle
 = {c_2Y_{\vec{0}}(x) Y_{\vec{0}}(y)\over \sqrt{\lambda_{eff}}}+\sum_{\vec{L}\neq 0}
 Y_{\vec{L}}(x) Y^*_{\vec{L}}(y)\left({1\over L(L+d)}+288\lambda_{eff}
 %{\Gamma(d+2/2)\over 2\pi^{(d+2)/2} }
 {c_4\over L^3(L+d)^3}\right)\label{sum3}
 \eea

\section{Comments on the in-in formalism}

We have demonstrated that calculations in the Euclidean theory yield finite
and calculable results in the massless limit. It is interesting to ask whether the
same statement is true for the in-in formalism for calculating correlation functions.
In this formalism, the correlation functions
at some time are defined by~\cite{weinberg1}
\bea
\langle Q(t)\rangle =
\int D\phi \exp(-i\int_{t_0}^t L(\phi_+) dt) Q(t) \exp(i\int_{t_0}^t L(\phi_-) dt)\d(\phi_+(t)-\phi_-(t))
\eea

We take $L(\phi)=\del\phi^2-m^2\phi^2-\lambda\phi^4$.
In perturbation theory, one  again finds that the $k$-loop diagram comes
 with a factor $({\lambda H^2\over m^2})^k$. This implies once again that the long wavelength modes are
 strongly coupled. (This fact has been known for a long time in the
 inflation literature~\cite{Starobinsky:1994bd,Vilenkin:1983xp}.)

In the case of the Euclidean sphere, there was only one mode (the $\vec{L}=0$) mode) that
was strongly coupled.
Here there are a continuum of modes with
small $k^2$
which are strongly coupled. This means
that we do not have an easily calculable theory; we still need to work in a strongly coupled
field theory as opposed to a theory of one mode. We will therefore be unable to
find exact results in this model, unlike the case of the Euclidean theory.

We can nevertheless make some qualitative statements for the correlation functions. We focus here on the
two point function in the massless theory. If
we are considering the two point function at short wavelengths and we  treat the
quartic interaction as a perturbation, we recover the
standard expression that
\bea
\langle \phi_k(t) \phi_k(t) \rangle \propto {1\over k^3}
\eea

On the other hand, for long wavelengths, we may expect the field to
develop a mass, just like the zero-mode on the Euclidean sphere.
 This mass will scale as some power of $\lambda$ i.e.  $m^2\propto \lambda^a$.
The $k$-loop diagram will then scale as
$({\lambda\over \lambda^a})^k $, and
will be small as along as the constant $a$ determining the scaling of the mass is small enough
i.e. if $a<1$. It would be interesting
to see if a nonperturbative approach can be used to calculate this scaling, and hence to justify
the validity of cosmological perturbation theory (a
 related but different approach has been suggested in~\cite{Burgess:2009bs}).

\section{Discussion and Conclusions}

We have considered an interacting scalar field in de Sitter space.
We showed that apparent infrared divergences which occur when the mass is small are the
result of an incorrect quantization of the zero-modes. In the Euclidean theory (which corresponds
to field theory on the Euclidean sphere), the
correct quantization
of the zero-mode leads to a finite and calculable perturbation expansion.
Unlike the massless noninteracting theory, where it is well known that no
de Sitter invariant vacuum exists, the interacting theory has a  vacuum with
unbroken de Sitter invariance.

Qualitatively, the inclusion of the interaction terms shifts
 the location in the pole of the zero-mode by an amount proportional to
 ${\sqrt{\lambda}}$.   This shift is physically
 different from a renormalization of the mass parameter,
 which can be taken to run to zero at low energies. The mass gap is
 purely due to the resummation
 of the quartic interactions; it is therefore analogous to the
 mass gap in QCD which can occur even
 for massless quarks.  In many ways, this result is similar to the
 analysis of the gap equation
 in~\cite{Riotto:2008mv}; however we have succeeded in obtaining the
 gap without the need for any approximation
because only one mode is strongly coupled.

We note that in some special theories,
several modes may be strongly coupled, which may make the theory uncalculable.
An example is the theory of a massless field coupled to a massive field with
$L=(\del\phi)^2+(\del\chi)^2-m^2\chi^2-\lambda\phi^2\chi^2$. Here,
the zero-mode of $\phi$ is strongly coupled to all the modes of $\chi$, and
the theory can be seen to be uncalculable. Nevertheless,
the
field theory does not break down, and correlation functions can be calculated numerically.

Finally, we note that we have been considering
a toy model with a scalar field on a fixed de Sitter background. In inflation, the metric is
not exactly de Sitter, which makes the Euclidean continuation problematic. It would be very
interesting to see whether our methods can be applied to this more realistic case.
We hope to return to this in  future work.

\section{Acknowledgements}
We would like to thank J.~Kumar and L.~Leblond for valuable discussions. This work is
supported in part by NSF Grant No. PHY-0653656.

 \end{document}